\documentclass[sigplan]{acmart}
\renewcommand\footnotetextcopyrightpermission[1]{} 
\usepackage{booktabs}   
\usepackage{listings}
\usepackage{enumitem}
\setlist{nolistsep,leftmargin=*}
\usepackage{subcaption}
\usepackage[T1]{fontenc}
\usepackage{microtype}
\usepackage{epstopdf}
\usepackage{flushend}
\usepackage{xspace}
\usepackage{paralist}

\newcommand{\systemname}[0]{Alchemy\xspace}
\newcommand{\pardis}[0]{SC\xspace}

\newcommand{\cut}[1]{}
\newcommand{\cutButInteresting}[1]{}

\newcommand{\TODOlater}[1]{}

\newcommand{\todoLATER}[1]{\textcolor{red}{\tiny{\bf [...]}}}

\newcommand{\codeColor}[0]{teal!90!black} 

\lstdefinelanguage{Scala}
{morekeywords={abstract,
  case,catch,class,
  def,
  else,extends,final,finally,for,
  if,import,implicit,
  match,macro,
  new,null,
  object,override,
  package,private,protected,
  return,
  super,sealed,
  this,throw,trait,try,type,
  val,var,lazy,
  with,while,
  yield,
  dsl,ir,q,code,bin,
  },
  sensitive,
  morecomment=[l]//,
  morecomment=[s]{/*}{*/},
  morestring=[b]`,
  showstringspaces=false
}[keywords,comments,strings]

\newcommand{\at}{@}

\lstMakeShortInline[columns=fixed,basicstyle=\footnotesize\ttfamily,keywordstyle=\scriptsize\ttfamily\bfseries,]@

\newcommand{\code}[1]{\lstinline[
  language=Scala,
  breaklines=true,
  showspaces=false,
  showtabs=false,
  showstringspaces=false,
  breakatwhitespace=true,
  numbers=none,
  numberstyle=\ttfamily,
  basicstyle=\ttfamily,
  keywordstyle=\ttfamily,%
  columns=fullflexible,
  escapeinside={(*@}{@*)}    
]|#1|}

\begin{document}

\title{A Compiler-Compiler for DSL Embedding}         

\author{Amir Shaikhha}
\affiliation{
  \institution{EPFL, Switzerland\\\{amir.shaikhha\}{\at}epfl.ch}
}

\author{Vojin Jovanovic}
\affiliation{
  \institution{Oracle Labs\\\{vojin.jovanovic\}{\at}oracle.com}
}

\author{Christoph Koch}
\affiliation{
  \institution{EPFL, Switzerland\\\{christoph.koch\}{\at}epfl.ch}
}

\begin{abstract}
In this paper, we present a framework to generate compilers for
embedded domain-specific languages (EDSLs).
This framework provides facilities to automatically generate the boilerplate code required for building DSL compilers on top of extensible optimizing compilers.
We evaluate the practicality of our framework by demonstrating several use-cases successfully built with it.

\end{abstract}



\maketitle

\section{Introduction}
\label{sec:intro}
\begin{quote}
{\em
Everything that happens once can never happen again. But everything that happens twice will surely happen a third time.
}\\
\hspace*{\fill}
-- Paulo Coelho, The Alchemist
\end{quote}

Domain-specific languages (DSLs) have gained enormous success in providing productivity and performance simultaneously. 
The former is achieved through their concise syntax, while the latter is achieved by using specialization and compilation techniques.
These two significantly improve DSL users' programming experience.

Building DSL compilers is a time-consuming and tedious task requiring much boilerplate code related to non-creative aspects of building a compiler, such as the definition of intermediate representation (IR) nodes and repetitive transformations~\cite{Book:1970:CSC:954344.954345}. 
There are many extensible optimizing compilers to help DSL developers by providing the required infrastructure for building compiler-based DSLs.
However, the existing optimizing compilation frameworks suffer from a steep learning curve, which hinders their adoption by DSL developers who lack compiler expertise.
In addition, if the API of the underlying extensible optimizing compiler changes, the DSL developer would need to globally refactor the code base of the DSL compiler.

The key contribution of this paper is to use a generative approach to help DSL developers with the process of building a DSL compiler.
Instead of asking the DSL developer to provide the boilerplate code snippets required for building a DSL compiler, we present a framework which automatically generates them.

More specifically, we present \systemname, a \textit{language workbench}~\cite{fowler2005language,erdweg2013state} for generating compilers for \textit{embedded DSLs} (EDSLs)~\cite{hudak-dsl} in the Scala programming language. 
DSL developers define a DSL as a normal library in Scala.
This plain Scala implementation can be used for debugging purposes without worrying about the performance aspects (handled separately by the DSL compiler).

\systemname provides a customizable set of annotations for encoding the domain knowledge in the optimizing compilation frameworks.
A DSL developer annotates the DSL library, from which \systemname generates a DSL compiler that is built on top of an extensible optimizing compiler.
As opposed to the existing compiler-compilers and language workbenches, \systemname does not need a new \textit{meta-language} for defining a DSL; instead, \systemname uses the reflection capabilities of Scala to treat the plain Scala code of the DSL library as the language specification.

A compiler expert can customize the behavior of the predefined set of annotations based on the features provided by a particular optimizing compiler.
Furthermore, the compiler expert can extend the set of annotations with additional ones for encoding various domain knowledge in an optimizing compiler.

This paper is organized as follows.
In Section~\ref{sec:background} we review the background material and related work. Then, in Section~\ref{sec:overview} we give a high-level overview of the \systemname framework. In Section~\ref{sec:metacompiler} we present the process of generating a DSL compiler in more detail. Section~\ref{sec:usecases} presents several use cases built with the \systemname framework. 
Finally, Section~\ref{sec:conclusions} concludes.

\section{Background \& Related Work}
\label{sec:background}
In this section, we present the background and related work to better understand the design decisions behind \systemname.

\subsection{Compiler-Compiler}
A compiler-compiler (or a meta compiler) is a program that generates a compiler from the specification of a programming language. 
This specification is usually expressed in a declarative language, called a \textit{meta-language}.

Yacc~\cite{johnson1975yacc} is a compiler-compiler for generating parsers specified using a declarative language.
There are numerous systems for
defining new languages, referred to as language workbenches~\cite{fowler2005language,erdweg2013state}, such as Stratego/Spoofax~\cite{spoofax}, SugarJ~\cite{Erdweg:2011:SLS:2048066.2048099}, Sugar*~\cite{Erdweg:2013:FEL:2517208.2517210}, KHEPERA~\cite{khepera}, and MPS~\cite{mps}.

\subsection{Domain-Specific Languages}
DSLs are programming languages tailored for a specific domain.
There are many successful examples of systems using DSLs in various domains such as SQL in database management, Spiral~\cite{spiral} for generating digital signal processing kernels, and Halide~\cite{ragan2013halide} for image processing.
The software development process can also be improved by using DSLs, referred to as language-oriented programming~\cite{lop}.
Cadelion~\cite{Lorenz:2011:CLL:2048066.2048123} is a language workbench developed for language-oriented programming.

There are two kinds of DSLs: 1) external DSLs which have a stand-alone compiler, and 2) embedded DSLs~\cite{hudak-dsl} (EDSLs) which are embedded in another generic-purpose programming language, called a \textit{host language}.

Various EDSLs have been successfully implemented in different host languages, such as Haskell~\cite{axelsson10:feldspar,qdsl,hudak-dsl} or Scala~\cite{lms,delite,Ofenbeck13:spiral-scala,Scherr:2015:AFL:2814204.2814217}.
The main advantage of EDSLs is reusing the existing infrastructure of the host language, such as the parser, the type checker, and the IDEs among others. 

There are two ways to define an EDSL. The first approach is by defining it as a plain library in the host language, referred to as \textit{shallowly} embedding it in the host language.
A shallow EDSL is reusing both the frontend and backend components of the host language compiler. 
However, the opportunities for domain-specific optimizations are left unexploited.
In other words, the library-based implementation of the EDSL in the host language is served an \textit{interpreter}.  

The second approach is \textit{deeply} embedding the DSL in the host language. 
A deep EDSL is only using the frontend of the host language, and
requires the DSL developer to implement a backend for the EDSL.
This way, the DSL developer can leverage domain-specific opportunities for optimizations and can leverage different target backends through code generation.

\subsection{Extensible Optimizing Compilers}
There are many extensible optimizing compilers which provide facilities for defining optimizations and code generation for new languages. 
Such frameworks can significantly simplify the development of the backend component of the compiler for a new programming language.

Stratego/Spoofax~\cite{spoofax} uses strategy-based
term-rewrite systems for defining domain-specific optimizations for
DSLs. Stratego uses an approach similar to quasi-quotation~\cite{stratego-qq}
to hide the expression terms from the user. For the same purpose, \systemname uses
annotations for specifying a subset of optimizations specified by the compiler expert. 
One can use quasi-quotes~\cite{scala-qq,Parreaux:2017:QSR:3136040.3136043} for implementing domain-specific optimizations in concrete syntax (rather than abstract syntax) similar to Stratego.

\subsection{What is \systemname?}
\systemname is a compiler-compiler, designed for EDSLs that use Scala as their host language.
\systemname uses the Scala language itself as its meta-language; it takes an annotated library as the implementation of a shallow EDSL
and produces the required boilerplate code for defining a backend for this EDSL using a particular extensible optimizing compiler.
In other words, \systemname converts an interpreter for a language (a shallow EDSL) to a compiler (a deep EDSL). 

Truffle~\cite{truffle-dsl} provides a DSL for defining
self-optimizing AST interpreters, using the Graal~\cite{graal} optimizing compiler as the backend. 
This system mainly focuses on providing just-in-time compilation for dynamically typed languages such as JavaScript and R, by annotating AST nodes.
In contrast, \systemname uses annotation on the library itself and generates the AST nodes based on strategy defined by the compiler expert.

Forge~\cite{forge} is an embedded DSL in Scala for specifying other DSLs. Forge
is used by the Delite~\cite{delite} and LMS~\cite{lms, rompf13popl} compilation
frameworks. This approach requires DSL developers to learn a new specification
language before implementing DSLs.  
In contrast, \systemname developers write a DSL specification using \textit{plain}
Scala code. Then, domain-specific knowledge is encoded using simple \systemname
annotations. 

Yin-Yang~\cite{yinyang} uses Scala macros for automatically converting
shallow EDSLs to the corresponding
deep EDSLs. 
Thus, it completely removes the need for providing the definition of a deep DSL
library from the DSL developer. However, contrary to our work, the compiler-compilation of Yin-Yang is specific to the LMS~\cite{lms} compilation framework. Also, Yin-Yang does not
generate any code related to optimizations of the DSL library. 
We have
identified the task of automatically generating the optimizations to be not only
a crucial requirement for DSL developers but also one that is significantly
more complicated than the one handled by Yin-Yang.

\section{Overview}
\label{sec:overview}
\begin{figure}[t!]
\centering
\includegraphics[width=\columnwidth]{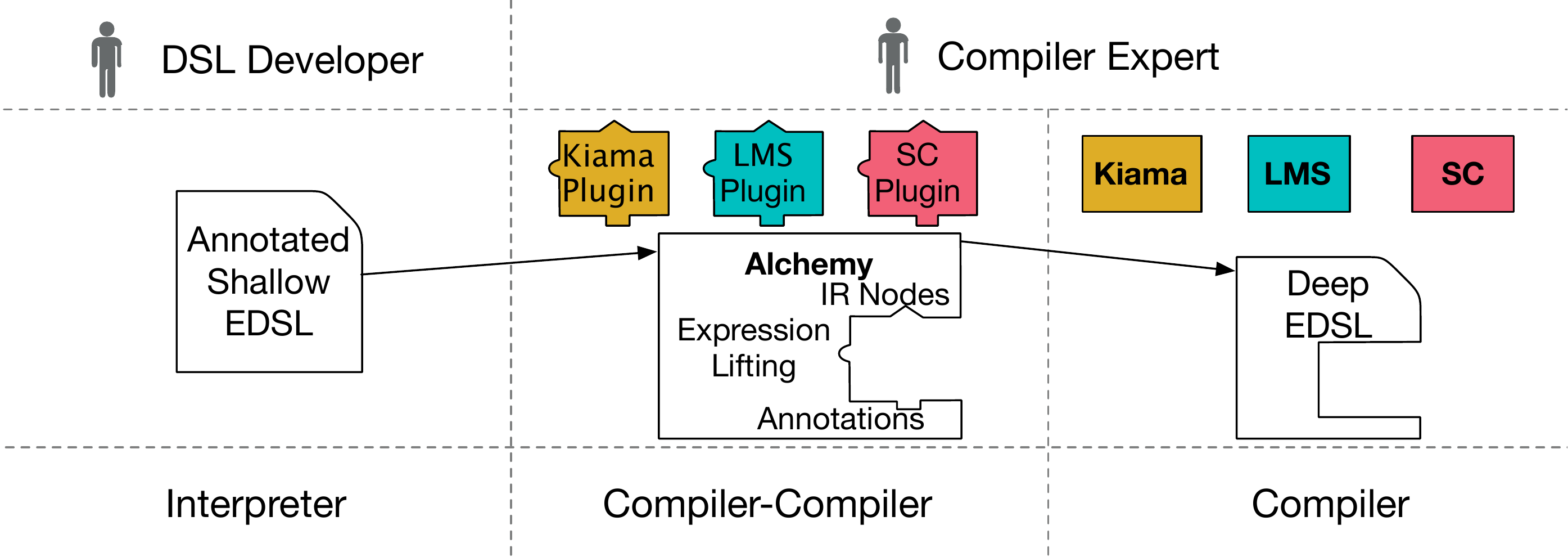}

\caption{Overall design of \systemname.}
\label{fig:systemdesign}
\end{figure}

Figure~\ref{fig:systemdesign} shows the overall design of the \systemname framework. 
\systemname is implemented as a compiler plugin for the Scala programming language.\footnote{We decided to implement \systemname as a compiler plugin rather than using the macro system of Scala, due to the restrictions imposed by def macros and macro annotations.}
After parsing and type checking the library-based implementation of an EDSL, \systemname uses the type-checked Scala AST to generate an appropriate DSL compiler. 
The generated DSL compiler follows the API provided by an extensible optimizing compiler to implement transformations and code generation needed for that DSL.

There are two different types of users for \systemname. 
The first type is a DSL developer, who is the end-user of the \systemname framework for defining a new DSL together with a set of domain-specific optimizations specified by a set of annotations. A DSL developer is a domain expert, without too much expertise in compilers. 

The second type of users is a compiler expert, who is not necessarily knowledgeable in various domains; instead, she is an expert in building optimizing compilers. 
In particular, a compiler expert has detailed knowledge about the internals of a specific extensible optimizing compiler. Thus, she can use the API provided by the \systemname framework to specify how the definition of an annotated Scala library is converted into the boilerplate code required for a DSL compiler built on top of an extensible optimizing compiler. 
Furthermore, she can extend the set of existing annotations provided by \systemname, for encoding the domain knowledge to be used by an optimizing compiler.

\section{Compiler-Compilation}
\label{sec:metacompiler}
In this section, we give more details on the process of generating a DSL compiler. 
First, we present the annotations defined by the \systemname framework. 
Then, we show the process of gathering the DSL information from an annotated library. 
Afterwards, through an example we give more details on the process of generating a DSL compiler based on the gathered DSL information. 
Then, we show how \systemname uses the implementation body of the annotated library for building DSL compilers. 
Finally, we show the process of generating EDSL compilers using a well-known embedding technique through our running example.

\subsection{\systemname Annotations}
\label{sec:annotations}

\noindent \textbf{Deep Types.} The DSL developers use the \code{\@deep} annotation for specifying the types for which they are interested in generating a corresponding deep embedding.
In other words, this annotation should be used for the types that are actually participating in the definition of a DSL, rather than helper classes which are used for debugging, profiling, and logging purposes.
\\

\noindent \textbf{Reflected Types.} The \code{\@reflect} annotation is used for
annotating the classes the source code of which the DSL developers have no access to.
This annotation is used in 
\systemname for
\begin{inparaenum}[a)]
\item annotating the methods of the Scala core libraries, such as HashMap, ArrayBuffer, etc.\ which are
frequently used, as well as for 
\item providing alternative implementations for the DSL library and the Scala core library. 
\end{inparaenum}
\\

\noindent \textbf{User-Defined Annotations.}
\systemname allows compiler experts
to define their custom annotations, together with the behavior of the target DSL
compiler for the annotated method. 
A compiler expert extends the API exposed by \systemname to implement the desired behavior (cf. Figure~\ref{fig:systemapi}).

\begin{figure}[t]
\centering
\begin{lstlisting}
case class ShallowDSL(types: List[ShallowType])
case class ShallowType(tpe: Type, 
  methods: List[ShallowMethod]) {
  def annotations: List[Annotation]
  def reflectType: Option[Type]
}
case class ShallowMethod(sym: MethodSymbol, 
  body: Option[Tree]) {
  def annotations: List[Annotation]
  def paramAnnots: List[(Int, Annotation)]
}
  
trait AlchemyCompiler {
  type DSLContext
  def liftType(t: Type)(implicit ctx: DSLContext): Type
  def liftExp(e: Tree)(implicit ctx: DSLContext): Tree
  def compileDSL(dsl: ShallowDSL)
        (implicit ctx: DSLContext): Tree
  def compileType(t: ShallowType)
        (implicit ctx: DSLContext): Tree
  def compileMethod(m: ShallowMethod)
        (implicit ctx: DSLContext): Tree
}
\end{lstlisting}

\caption{The API of \systemname for compiler experts.}
\label{fig:systemapi}
\end{figure}

\subsection{Gathering DSL Information}
The \systemname framework inspects the Scala AST of the given annotated library after the type checking phase of the Scala compiler.
Based on the typed Scala AST, \systemname produces 
the information about the shallow version of the EDSL by building \code{ShallowMethod}, \code{ShallowType}, and \code{ShallowDSL} objects, corresponding to the DSL methods, DSL types, and the whole DSL, respectively.

A \code{ShallowMethod} instance has the symbol of the DSL method (the \code{sym} parameter) and the AST of its body, if available.
Also, this instance returns the list of annotations that the DSL developer has used for the method (\code{annotations}) and its parameters (\code{paramAnnots}).

A \code{ShallowType} instance contains the information of the DSL type (the \code{tpe} parameter) and the list of its methods.
In addition, this instance has the list of annotations used for the type (\code{annotations}) and the type it reflects (\code{reflectType}) in the case where it is annotated with \code{\@reflect}.

Finally, a \code{ShallowDSL} instance has the information of all DSL types that are annotated with \code{\@deep}. Next, we show how this information is used to build a compiler for a simple DSL. 

\subsection{Generating an EDSL Compiler}
Let us consider a DSL for working with complex numbers as our running example. 
For this DSL, we generate a DSL compiler using a simple form of expression terms as the intermediate representation, which is used by compilation frameworks such as Kiama~\cite{kiama}. 

Figure~\ref{fig:complexshallow} shows the implementation of this EDSL as an annotated Scala library.
This implementation can be used as a normal Scala library to  benefit from all the tool-chains provided for Scala such as debugging tools and IDEs.

\begin{figure}[t]
\centering
\begin{lstlisting}
@deep
class Complex(val re: Double, val im: Double)
object Complex {
  def add(c1: Complex, c2: Complex): Complex = 
    new Complex(c1.re + c2.re, c1.im + c2.im)
  def sub(c1: Complex, c2: Complex): Complex = 
    new Complex(c1.re - c2.re, c1.im - c2.im)
  def zero(): Complex =
    new Complex(0, 0)
}
\end{lstlisting}

\caption{The annotated complex DSL implementation.}
\label{fig:complexshallow}
\end{figure}

Figure~\ref{fig:complexdeep} shows the definition of IR nodes generated by \systemname.
The IR nodes are algebraic data types (ADTs), each one specifying a different
construct of the Complex DSL. 
For each method of the \code{Complex} companion object, \systemname generates a case class
with a default naming scheme in which the name of the object is followed by the name of the method. For example, the method \code{add} of the \code{Complex} object is converted to the \code{ComplexAdd} case class.
As another example, the constructor of the \code{Complex} class is converted to the \code{ComplexNew} case class.
Each case class has the same number of arguments as the corresponding shallow method.

\begin{figure}[t]
\centering
\begin{lstlisting}
// Predefined by a compiler expert
trait Exp
case class DoubleConstant(v: Double) extends Exp
// Automatically generated by Alchemy
case class ComplexNew(re: Exp, im: Exp) extends Exp
case class ComplexAdd(c1: Exp, c2: Exp) extends Exp
case class ComplexSub(c1: Exp, c2: Exp) extends Exp
case class ComplexZero() extends Exp
\end{lstlisting}

\caption{The generated IR nodes for the Complex DSL.}
\label{fig:complexdeep}
\end{figure}

The methods of a class are converted in a similar manner. 
The key difference is that the generated case class has an additional argument corresponding to \code{this} object. 
For example, the method \code{+} of the \code{Complex} class is converted to a case class with two parameters, where the first parameter corresponds to \code{this} object of the \code{Complex} class, and the second parameter corresponds to the input parameter of the \code{+} method.

As explained before, \systemname allows a compiler expert to define user-defined annotations. 
In Figure~\ref{fig:complexshallow2}, the \code{\@name} annotation is used for overriding the default naming scheme provided by \systemname. 
For example, the \code{+} method is converted to the \code{ComplexAdd} case class.

\begin{figure}[t]
\centering
\begin{lstlisting}
@deep
class Complex(val re: Double, val im: Double) {
  @name("ComplexAdd")
  def +(c2: Complex): Complex = 
    new Complex(this.re + c2.re, this.im + c2.im)
  @name("ComplexSub")
  def -(c2: Complex): Complex = 
    new Complex(this.re - c2.re, this.im - c2.im)
}
object Complex {
  def zero(): Complex =
    new Complex(0, 0)
}
\end{lstlisting}

\caption{The second version of the annotated Complex DSL implementation.}
\label{fig:complexshallow2}
\end{figure}

\subsection{Lifting the Implementation}
As Figure~\ref{fig:systemapi} shows, \systemname also provides two methods for lifting the expression and the type of the implementation body of DSL library methods.
These two methods are useful for defining syntactic sugar constructs for a DSL (i.e., the DSL constructs that do not have an actual node in the compiler, instead they are defined in terms of other constructs of the DSL). 
An example of such a construct can be found in Section~\ref{sec:compcomp_poly}.

In addition, by providing several reflected versions (cf. Section~\ref{sec:annotations}) for a particular type, each one with a different implementation, \systemname can generate 
several transformations for those DSL constructs.
This removes the need to implement a DSL IR transformer, which manipulates the IR defined in the underlying optimizing compiler.

To specify the way that expressions should be transformed, compiler experts can implement a Scala AST to Scala AST transformation (cf. the \code{liftExp} method in Figure~\ref{fig:systemapi}).
Note that implementing Scala AST to Scala AST transformations from scratch can be a tedious and time-consuming task. Alternatively, if the target optimizing compiler uses the tagless final~\cite{finally} or polymorphic embedding~\cite{polymorphic} approaches, one can use frameworks such as Yin-Yang~\cite{yinyang}, which are already providing the translation required for these approaches. 
Next, we show a DSL compiler generated based on the polymorphic embedding approach.

\begin{figure}[t]
\centering
\begin{lstlisting}
// Shallow expression
new Complex(2, 3) - Complex.zero()
// Lifted expression
ComplexSub(
  ComplexNew(
    DoubleConstant(2), DoubleConstant(3)
  ), ComplexZero()
)
\end{lstlisting}

\caption{An example expression and its lifted version in Complex DSL.}
\label{fig:liftshallow}
\end{figure}

\subsection{Generating a Polymorphic EDSL Compiler}
\label{sec:compcomp_poly}

Let us consider the third version of the Complex DSL, shown in Figure~\ref{fig:complexshallow3}. 
This version has an additional construct for negating a complex number, specified by the \code{unary\_-} method.  
Subtracting two complex numbers is a syntactic sugar (annotated with \code{\@sugar}) for adding the first complex number with the negation of the second complex number.

\begin{figure}[t]
\centering
\begin{lstlisting}
@deep
class Complex(val re: Double, val im: Double) {
  @name("ComplexAdd")
  def +(c2: Complex): Complex = 
    new Complex(this.re + c2.re, this.im + c2.im)
  @name("ComplexNeg")
  def unary_-(): Complex =
    new Complex(-this.re, -this.im)
  @sugar
  def -(c2: Complex): Complex = 
    this + (-c2)
}
object Complex {
  def zero(): Complex =
    new Complex(0, 0)
}
\end{lstlisting}

\caption{The third version of the annotated Complex DSL implementation.}
\label{fig:complexshallow3}
\end{figure}

Polymorphic embedding~\cite{polymorphic} (or tagless final~\cite{finally}), is an approach for implementing EDSLs where every DSL construct is converted into a function (rather than an ADT) and the interpretation of these functions are left abstract. 
Thus, it is possible to provide such abstract interpretations with different instances, such as actual evaluation, compilation, and partial evaluation ~\cite{polymorphic,finally}. 

Figure~\ref{fig:complexpolyinterface} shows the polymorphic embedding interface generated by \systemname for the third version of the Complex DSL.
The type member \code{Rep[T]} is an abstract type representation for different interpretations of Complex DSL programs.

\begin{figure}[t]
\centering
\begin{lstlisting}
trait ComplexOps {
  type Rep[T]
  def doubleConst(d: Double): Rep[Double]
  def complexAdd(self: Rep[Complex], 
        c2: Rep[Complex]): Rep[Complex]
  def complexNeg(self: Rep[Complex]): Rep[Complex]
  def complexSub(self: Rep[Complex], 
        c2: Rep[Complex]): Rep[Complex]
  def complexZero(): Rep[Complex]
  def complexNew(re: Rep[Double], 
        im: Rep[Double]): Rep[Complex]
}
\end{lstlisting}
\caption{The generated polymorphic embedding interface for the Complex DSL.}
\label{fig:complexpolyinterface}
\end{figure}

Figure~\ref{fig:complexpolydeep} shows the generated deep embedding interface for the polymorphic embedding of the Complex DSL. 
Instead of using ADTs for defining IR nodes, this time we use generalized algebraic data types (GADTs).
The invocation of each DSL construct method results in the creation of the corresponding node.
As the subtraction of two complex numbers is a syntactic sugar, no corresponding IR node is created for it. 
Instead, the \code{complexSub} method results in the invocation of the \code{complexAdd} and \code{complexNeg} methods, which is generated using the \code{liftExp} method of \systemname.

\begin{figure}[t]
\centering
\begin{lstlisting}
// Predefined by a compiler expert
trait Exp[T]
case class DoubleConstant(d: Double) extends Exp[Double]
// Automatically generated by Alchemy
case class ComplexNew(re: Exp[Double], 
                      im: Exp[Double]) extends Exp[Complex]
case class ComplexAdd(self: Exp[Complex], 
                      c2: Exp[Complex]) extends Exp[Complex]
case class ComplexNeg(self:Exp[Complex])extends Exp[Complex]
case class ComplexZero() extends Exp[Complex]

trait ComplexExp extends ComplexOps {
  type Rep[T] = Exp[T]
  def doubleConst(d: Double): Rep[Double] =
    DoubleConstant(d)
  def complexAdd(self: Rep[Complex], 
        c2: Rep[Complex]): Rep[Complex] =
    ComplexAdd(self, c2)
  def complexNeg(self: Rep[Complex]): Rep[Complex] =
    ComplexNeg(self)
  def complexSub(self: Rep[Complex], 
        c2: Rep[Complex]): Rep[Complex] = 
    complexAdd(self, complexNeg(c2))
  def complexZero(): Rep[Complex] = 
    ComplexZero()
  def complexNew(re: Rep[Double], 
        im: Rep[Double]): Rep[Complex] = 
    ComplexNew(re, im)
}
\end{lstlisting}
\caption{The generated IR node definitions and deep embedding interface for the Complex DSL.}
\label{fig:complexpolydeep}
\end{figure}

Figure~\ref{fig:liftshallow3} shows the lifted expression of the example of Figure~\ref{fig:liftshallow}.
In this case, instead of converting expressions to their ADT definition, \systemname converts them to their corresponding DSL method definition in polymorphic embedding.
In addition, this figure shows the generated IR nodes for this program, in which the subtraction construct is desugared into the addition and negation nodes. 
Note that the negation of zero and addition with zero can be further simplified by providing yet another optimized interface implementation in polymorphic embedding. 
Examples of such simplifications are given later in Section~\ref{sec:generating-online-transf}.

\begin{figure}[t]
\centering
\begin{lstlisting}
// lifted expression in polymorphic embedding
complexSub(
  complexNew(
    doubleConst(2), doubleConst(3)
  ), complexZero()
)
// generated IR nodes
ComplexAdd(
  ComplexNew(
    DoubleConstant(2), DoubleConstant(3)
  ), ComplexNeg(
    ComplexZero()
  )
)
\end{lstlisting}

\caption{Polymorphic embedding version of the example in Figure~\ref{fig:liftshallow}, and the generated IR nodes.}
\label{fig:liftshallow3}
\end{figure}

Up to now, we have used simple expression terms for the definition of IR nodes.
\systemname can easily generate other types of IR nodes such as A-Normal Form~\cite{anf},
where the children of a node are either constant values or variable accesses. This means that
all non-trivial sub-expressions are let-bound, which helps in applying optimizations such as common-subexpression elimination (CSE) and dead-code elimination (CSE).
Such normalized types of IR nodes are used in various optimizing compilers such as Graal~\cite{graal}, LMS~\cite{lms}, Squid~\cite{Parreaux:2017:QSR:3136040.3136043}, and \pardis.
We will see more detailed examples of \pardis in the next section.

\section{Use Cases and Evaluation}
\label{sec:usecases}
In this section, we present the use cases built on top of \systemname. We provide an extended set of annotations and show their usage. 
Finally, we evaluate the productivity of the DSL developer.

\subsection{\pardis}

\begin{figure}[t!]
\centering
\includegraphics[width=\columnwidth]{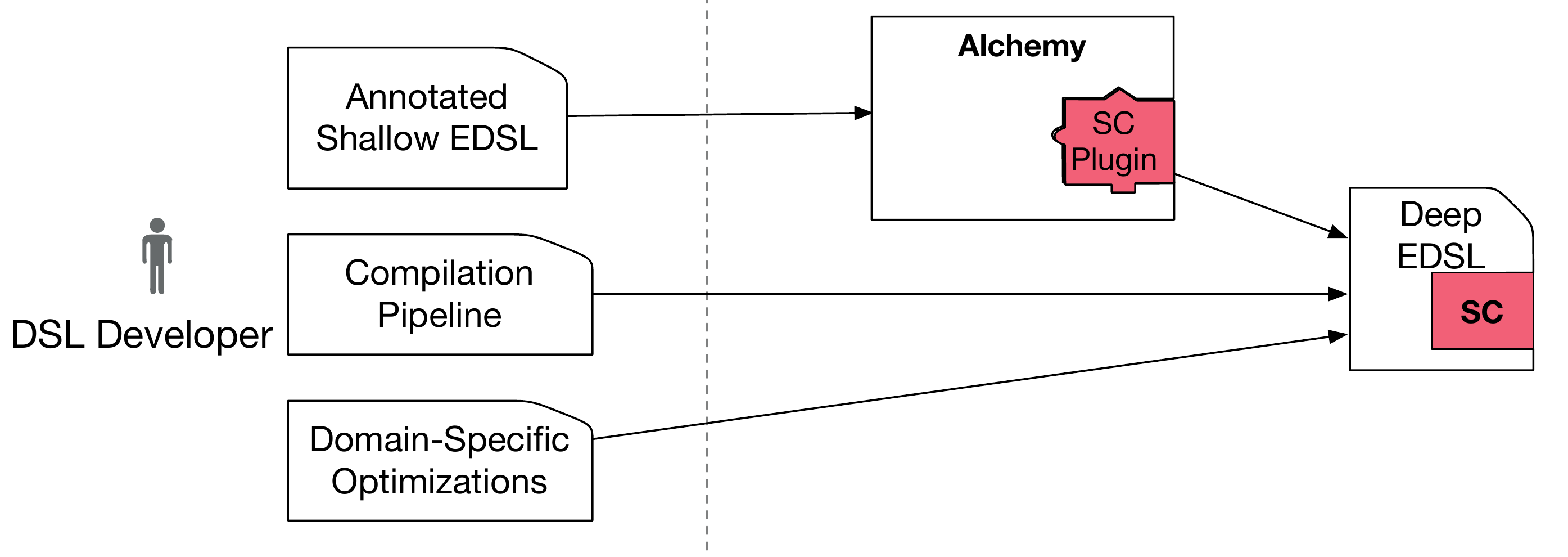}

\caption{Overall design of \pardis used with \systemname.}
\label{fig:work-flow-perf}
\end{figure}

\pardis (the Systems Compiler) is a compilation framework for building compilation-based systems in the Scala programming language. 
Different system component libraries can be considered as different DSLs, for which system developers extend \pardis to build DSL compilers.
To hide the internal implementation details of the compiler, \systemname provides an abstraction layer between the system component libraries and
the \pardis optimizing compiler itself. 
Figure~\ref{fig:work-flow-perf} shows the
overall design of \systemname and \pardis, which operates as follows.


The system developer (who is actually a DSL developer) uses the \pardis plugin of \systemname to create a DSL compiler.
Many systems optimizations are automatically converted by \systemname 
to functions that manipulate the IR of the compiler. The system developer
uses a set of \textit{annotations} provided by the compiler expert of the \pardis framework, to specify the IR 
transformations. 
To provide more advanced domain-specific optimizations that cannot be encoded by annotations, as well as compilation phases, the system developer uses the transformation API provided by \pardis.

\pardis converts the systems code to a graph-like intermediate 
representation (IR). As \pardis follows the polymorphic embedding approach~\cite{polymorphic} for deeply embedding DSLs, \pardis uses
Yin-Yang~\cite{yinyang}\footnote{We note that Yin-Yang, in contrast to our work,
handles only the conversion from plain Scala code to IR, without providing any
functionality related to code optimization of the systems library.} which applies several
transformations (e.g., language
virtualization~\cite{Chafi:2010:LVH:1932682.1869527}) in order to convert the
plain Scala code into the corresponding IR.

We have used \pardis to build two different compilation-based query engines: \begin{inparaenum}[a)] \item an analytical query processing engine~\cite{legobase_tods,dblablb}, and \item a transactional query processing engine~\cite{dashti_beta}.
\end{inparaenum}

From the perspective of the abstraction level of a program, the transformations
are classified into two categories. First, \textit{optimizing} transformations
transform a program into another program on the \textit{same} level of abstraction.
Second, \textit{lowering} transformations convert a program into one on a
\textit{lower} abstraction level. 
\pardis provides a set of built-in transformations out-of-the-box.  These mainly
consist of generic compiler optimizations such as common-subexpression
elimination (CSE), dead-code elimination (DCE), partial evaluation (PE), etc. 

The last phase in
the \pardis compiler is code generation where the compiler generates the code
based on the desired target language.  Observe that since each lowering
transformation brings the program closer to the final target code, this provides the
excellent property that code generation (e.g., C code generation) in the end basically becomes a
trivial and na\"ive stringification of the lowest level representation.

For converting from host to target languages, \pardis can make use of the
same infrastructure.  To do this
conversion, a DSL developer only has to express the constructs and the
necessary data-structure API of the target language as a library inside the host
language. 
Then, there 
is no need for the DSL
developer to \textit{manually} provide code generation for the target language using internal
compilers APIs as is the case with most existing solutions.
In contrast, \systemname automatically generates the transformation phases needed to
convert from host language IR nodes to target language IR nodes (e.g., from
Scala to C). 

An important side-effect of our design is that since the plain Scala code of a
system does not require \textit{any} specific syntax, type or IR-related
information from \pardis, this code is \textit{directly} executable using
the normal Scala compiler.
In this case, the Scala compiler will ignore all \systemname
annotations, and interpret the code of the system using plain Scala.  \systemname
can thus be seen as a system for converting a \textit{system interpreter}
(which executes the systems code unoptimized) into the corresponding
\textit{system compiler} along with its \textit{optimizations}.

\begin{figure}[t]
\centering
\begin{lstlisting}
class MyTransformer extends RuleBasedTransformer {
  analysis += rule { case Pattern =>
    // gather necessary information for analysis
  }
  rewrite += rule { case Pattern =>
    /* use analysis information while generating
       the appropriate transformed node */
  }
  rewrite += remove { case Pattern =>
    /* use analysis information to remove node */ 
  }
}
\end{lstlisting}    

\vspace{-3mm}

\caption{Offline Transformation API of \pardis.}
\label{fig:transformation-api}
\end{figure}
 
Next, we briefly provide more details about the two categories of transformations
that \pardis supports.

\subsection{\pardis Transformations} 
\label{sec:pardis}
\pardis classifies the transformations into two categories, which we present in
more detail next while also highlighting differences from previous work in each
class.  
\\

\noindent\textbf{Online transformations} are applied while the IR nodes are
generated. Every construct of a DSL is mapped to a method invocation, which in
turn results in the generation of an IR node~\cite{finally, polymorphic}. By
overriding the behavior of that method, an online transformation can result in
the generation of a different (set of) IR node(s) than the original IR node.
Even though a large set of optimizations (such as constant folding, common
subexpression elimination, and others) can be expressed using online
transformations, some optimizations need to be preceded by analysis over the whole program. 

For a restricted set of control-flow constructs, namely \textit{structured loops},
it is possible to use the Speculative Rewriting~\cite{speculative} approach in
order to combine the data-flow analysis with an online transformation, thus
bypassing the need for a separate analysis pass. However, we
have observed that there exists an important class of transformations in which
the corresponding analysis \textit{cannot} be combined with the transformation
phase.  This class of optimizations, which cannot be handled by existing
extensible optimizing compilers, is presented next.
\\

\noindent\textbf{Offline transformations} need whole program analysis
before applying any transformation. Figure~\ref{fig:transformation-api} shows
the \pardis offline transformation API.  The \code{analysis} construct
specifies the information that should be collected during the analysis phase of
a transformation. The \code{rewrite} construct specifies the transformation
rules based on the information gathered during the analysis phase. Finally, the
\code{remove} construct removes the pattern specified in its body. 

The \systemname annotation processor takes care of converting the Scala
annotations of the systems library, which express optimizations, into IR
transformers which manipulate the intermediate representation of \pardis. 
This is explained in more detail in Section~\ref{sec:generating-online-transf}.

\subsection{\pardis Annotations}
\label{subsec:annotations}
In this section, we present in more detail the different categories of
annotations implemented for \pardis.

\noindent 
\textbf{Side-Effects.} These are annotations that guide the effect system of the
optimizing compiler. For example, a method annotated with \code{\@pure} denotes
that this method does not cause any side effects and the expressions that call this method can be moved freely throughout the program. In
addition, \systemname provides more fine-grained effect annotations that keep track
of read and write mutations of objects. More precisely, if a method is annotated
with \code{read} or \code{write} annotations, then there exists a mutation
effect over the specific object (i.e.,\ \code{this}) of that particular class.
Similarly, an annotated argument may include read or write effects over
that argument.
\begin{figure}[t]
\centering
\begin{lstlisting}
@deep (*@@*)
@inline (*@@*)
abstract class Operator[A] {
  abstract def init(): Unit
}

@deep (*@@*)
@inline (*@@*)
class ScanOp[A](table: Array[A]) extends Operator[A]{
  var i = 0
  @inline (*@@*)
  def init() = {
    while (i < table.length) {
      child.consume(table(i))
      i += 1
    }
  }
}

@deep (*@@*)
@inline (*@@*)
class HashJoinOp[A,B,C](val leftParent: Operator[A], 
  val rightParent: Operator[B], ...) extends
    Operator[CompositeRecord[A, B]] {
  @inline var mode = 0
  @inline (*@@*)
  def init() = {
    mode = 0
    // phase 1: leftParent will call this.consume
    leftParent.init()
    mode = 1
    // phase 2: rightParent will call this.consume
    rightParent.init()
    mode = 2
  }
  @inline (*@@*)
  def consume(tuple: Record): Unit = {
    if (mode == 0) {
      /*phase1 -- elided code for left side of join*/
    } else if (mode == 1) {
      /*phase2 -- elided code for right side of join*/
    }
  }
}
\end{lstlisting}

\vspace{-3mm}

\caption{Inline annotations of two operators in our analytical query engine.}
\label{fig:operator-inline-annotation}
\end{figure}
\\

\noindent \textbf{Inline.} The \code{\@inline} annotation guides the inlining
decisions of the compiler.  This annotation can be applied to methods, whole
classes as well as class fields, with different semantics in each case.
Methods annotated with the \code{\@inline} annotation specify that every
invocation of that method should be inlined by the compiler. For classes, the
\code{\@inline} annotation removes the abstraction of the specific
class during compilation time. In essence, this means that the methods of an
inlined class are implicitly annotated with the inline annotation and are
subsequently inlined. This makes inlined classes in \systemname semantically
similar to value classes~\cite{value-class-java}.
Finally, a mutable field of a class can also be annotated with \code{\@inline},
which means that all the usages of this field are partially evaluated during compilation time. 

Figure~\ref{fig:operator-inline-annotation} shows the scan and hash-join operators
annotated with the \code{\@inline} annotation. 
In this example, all methods of the \code{HashJoinOp} class are automatically
inlined, as the \code{HashJoinOp} class is marked with the \code{\@inline}
annotation. Furthermore, the mutable field \code{mode} is partially evaluated
at compilation time 
and, as a result, the corresponding branch in the \code{consume} method is also
partially evaluated at compilation time. More concretely, both
\code{leftParent.init} and \code{rightParent.init} invoke the \code{consume}
method of the \code{HashJoinOp} class. 
However, the former inlines the code in the
\code{phase1} block whereas the latter inlines the \code{phase2} code block.
This is possible as \code{mode} is evaluated during compilation time and, thus,
there is no need to generate any code for it and the corresponding
\code{if} condition checks. We have found that there are multiple examples where
such \code{if} conditions can be safely removed in our analytical query engine (e.g.,\ in the case
of configuration variables whose values are known in advance at startup time). 
\\

\noindent \textbf{Algebraic Structure.} These are annotations for specifying the
common algebraic rules that occur frequently for different use cases. For
example, \code{\@monoid} specifies a binary operation of a type that has a monoid structure.  In the case of natural numbers, \code{\@monoid(0)} over the \code{+}
operator represents that \code{a+0=0+a=a}.  The annotation processor generates
several constant folding optimizations which benefit from such algebraic
structure and significantly improve the performance of systems that use them. 

Furthermore, the \code{\@commutative} annotation specifies that the order of the
operands of a binary operation can be changed without affecting the result.
This property is useful for applying constant folding on cases in which static
arguments and dynamic arguments are mixed in an arbitrary order, thus hindering
the constant folding process. For example, in the expression \code{1 + a + 2},
constant folding cannot be performed without specifying that the commutativity
property of addition on natural numbers is applicable in this case. However,
if we push the static terms to the left side of the expression while we generate
the nodes, we generate the IR which represents the expression \code{1 + 2 + a}
instead of the previous expression. Then, it becomes possible to apply constant
folding and get the expression \code{3 + a}.
\\

\subsection{Generating Transformation Passes}
\label{sec:generating-online-transf}
As discussed in
Section~\ref{sec:pardis}, these transformation passes are classified into two
categories: \textit{online} and \textit{offline} transformations.
In this section, we demonstrate how \systemname generates online and offline transformation passes.
\\

\noindent \textbf{Generating Online Transformations.}
In general,
\systemname uses node generation (online transformation) in order to implement
the appropriate rewrite rules for most annotations.
As we discussed in 
Section~\ref{sec:compcomp_poly}, every construct of a DSL is mapped to a method invocation, which in
turn results in the generation of an IR node~\cite{finally, polymorphic}.

For example, in the case of addition on natural numbers, the default behavior
for the method \code{int\_plus} is shown in lines 1-3
of Figure~\ref{fig:online-transformation}. This method generates the
\code{IntPlus} IR node, which is also automatically generated by \systemname.
However, when this method is annotated with the \code{\@monoid} and
\code{\@commutative} annotations, this results in the generation of an online
transformation. More specifically, the annotated method automatically generates
the code shown in lines 5-14 of the same figure. First, as the method is pure,
\pardis checks if both arguments are statically known. This is achieved
by checking if the expressions are of \code{Constant} type or not. In this case,
\pardis performs partial evaluation by computing the result through the addition of the 
arguments. Second, if only one
of the arguments is statically known and it is equal to 0, the monoid property
of this operator returns the dynamic operand. Third, if only one of the
arguments is statically known (but it is not zero), then the static argument
is pushed as the left operand, as we know that this operator is commutative. 
Finally, if none of the previous cases is true, then the default behavior is
used and the original \code{IntPlus} IR node is generated.

\systemname also generates an online transformation out of the \code{\@inline}
annotation. For methods with this annotation, instead of generating the
corresponding node, \systemname generates the nodes for the body of that method.
In the special case of dynamic dispatch, the concrete type of the object is
looked up and based on its value \systemname invokes the appropriate method.
\begin{figure}[t]
\centering
\begin{lstlisting}
@deep (*@@*)
@reflect[Int] (*@@*)
class AnnotatedInt {
  @commutative 
  @monoid(0) 
  @pure 
  def +(x: Int): Int
}
\end{lstlisting}

\caption[\systemname annotations of the \code{Int} class.]{\systemname annotations of the \code{Int} class. The \code{AnnotatedInt}
class is a mirror class for the original \code{Int} Scala class.}
\label{fig:monoid-int}
\end{figure}
\begin{figure}[t]
\centering
\begin{lstlisting}
trait Base {
  type Rep[T]
}

trait BaseExp extends Base {
  type Rep[T] = Exp[T]
}

trait IntOps extends Base {
  def int_plus(a: Rep[Int], b: Rep[Int]): Rep[Int]
}

trait IntExp extends IntOps with BaseExp {
  // default IR generation
  def int_plus(a: Exp[Int], b: Exp[Int]): Exp[Int] = 
    IntPlus(a, b)
}

trait IntExpOpt extends IntExp {
  // optimized IR generation
  override 
  def int_plus(a: Exp[Int], b: Exp[Int]): Exp[Int] = 
  (a, b) match {
    case (Constant(aStatic), Constant(bStatic)) => 
      Constant(aStatic + bStatic)
    case (Constant(0), bDynamic) => bDynamic
    case (aDynamic, Constant(0)) => aDynamic
    case (aDynamic,Constant(bStatic)) => int_plus(b,a)
    case (_, _) => super.int_plus(a,b)
  }
}
\end{lstlisting}

\caption{The generated online transformation by \systemname for addition on 
\code{Int}.}
\label{fig:online-transformation}
\end{figure}

For example, the annotated code for the scanning operator of the analytical query engine, shown in
Figure~\ref{fig:operator-inline-annotation}, generates the compiler code shown in
Figure~\ref{fig:legobase-inline}.
There the \code{scanOpInit} method represents the corresponding method
which is invoked in order to generate an appropriate IR. As is the case with
integer addition, the default behavior of this method, which results in
creating the \code{ScanOpInit} IR node, is shown in lines 1-3. The rest of the
code presents the implementation of the \code{\@inline} annotation for this
operator, which results in inlining the body of this method while generating the IR node. The method \code{scanOpInit} is automatically generated by
\systemname which generates the body of the \code{init} method. As described earlier,
all method invocations lead to the generation of the corresponding IR nodes.
For example, \code{\_\_whileDo} results in creating an IR node for a \code{while} loop.
Finally, for inlining the \code{init} method of the \code{Operator} class, we need
to handle dynamic dispatch, as we described earlier. 
We do so by redirecting to the appropriate method
based on the type of the caller object.
An alternative design is to use multi-stage programming for encoding the fact that the objects of Operator class are staged away.
This is achieved by generating the deep embedding interface of all operator classes as partially static. 
With a similar design, one can support staging for other libraries implemented using design patterns that require abstraction overheads such as generic programming~\cite{Lammel:2005:SYB:1086365.1086391,yallop2017staged}.

\begin{figure}[t]
\centering
\begin{lstlisting}
trait OperatorOps extends Base {
  def operatorInit[A:Type](self:Rep[Operator[A]]):Rep[Unit]
}

trait ScanOpOps extends OperatorOps {
  def scanOpInit[A:Type](self: Rep[ScanOp[A]]): Rep[Unit] 
}

trait OperatorExp extends OperatorOps with BaseExp {
  def operatorInit[A:Type](self: Exp[Operator[A]]) = 
    OperatorInit(self)
}

trait ScanOpExp extends ScanOpOps with OperatorExp {
  // the default behavior of scanOp.init operation
  def scanOpInit[A:Type](self: Exp[ScanOp[A]]) = 
    ScanOpInit(self)
}

trait ScanOpInline extends ScanOpExp {
  // the inlined behavior of scanOp.init operation
  override
  def scanOpInit[A:Type](self: Exp[ScanOp[A]]) = 
    __whileDo(self.i < (self.table.length), {
      self.child.consume(self.table.apply(self.i))
      self.i = self.i + unit(1)
    })
  // handling of dynamic dispatch for operator.init
  override
  def operatorInit[A:Type](self: Exp[Operator[A]]) = 
    self.tpe match {
      case ScanOpType(_) =>
        scanOpInit(self.asInstanceOf[Exp[ScanOp[A]]])
      // the rest of the operator types ...
    }
}
\end{lstlisting}
\caption{The generated online transformations by \systemname for the scan
operator of the analytical query engine.}
\label{fig:legobase-inline}
\end{figure}

\begin{figure*}[t]
\begin{subfigure}[t]{0.33\textwidth}
\begin{lstlisting}[numbers=none]
@offline
@reflect[Seq[_]] (*@@*)
class SeqLinkedList[T] {
  var head: Cont[T] = null

  
  def +=(elem: T) = 
    head = Cont(elem, head)
  
  
  def foreach(f: T => Unit) = {
    var current = head
    while (current != null) {
      f(current.elem)
      current = current.next
    }
  }
}
\end{lstlisting}
\end{subfigure}
\begin{subfigure}[t]{0.33\textwidth}
\begin{lstlisting}[numbers=none]
@offline
@reflect[Seq[_]] (*@@*)
class SeqArray[T: Manifest] {
  val array = 
    new Array[T](MAX_BUCKETS)
  var size: Int = 0
  def +=(elem: T) = {
    array(size) = elem
    size += 1
  }
  def foreach(f: T=>Unit) =
    for (i <- 0 until size) {
      val elem = array(i)
      f(elem)
    }


}
\end{lstlisting}
\end{subfigure}
\begin{subfigure}[t]{0.33\textwidth}
\begin{lstlisting}[numbers=none]
@offline
@reflect[Seq[_]] (*@@*)
class SeqGlib[T] {
  var gHead: Pointer[GList[T]] = null
  
  
  def +=(x: T) = 
    gHead = 
      g_list_append(gHead, &(x))
    
  def foreach(f: T=>Unit) = {
    var current = gHead
    while (current != NULL) {
      f(*(current.data))
      current = g_list_next(current)
    }
  }
}
\end{lstlisting}
\end{subfigure}
\caption{Different transformations for the Scala Seq class. 
The transformations are written using plain Scala code.}
\label{fig:seqlowering}
\end{figure*}

\noindent \textbf{Generating Offline Transformations.}
The generated transformations are not limited to online transformations.
\systemname also generates offline transformation passes. 
Figure~\ref{fig:seqlowering} shows the implementation of three different transformations for the \code{Seq} class\footnote{By a \code{Seq} data type, we mean a collection where the order of its elements does not matter.}, in plain Scala code. The first implementation uses a linked list for storing the elements of the sequence. 
The second implementation stores the elements in an array data-structure.\footnote{This implementation assumes that the number of the elements in the collection does not exceed \code{MAX_BUCKETS}. 
In cases where this assumption does not hold, one has to make the corresponding field mutable, and add an additional check while inserting an element.} 
Finally, the third implementation uses a \code{g_list} data-structure, provided by GLib.
The generated transformation from this class can be used for using data structures provided by GLib in the generated C code.

These implementations can be used for debugging the correctness of the transformers.
For using them in the DSL compiler, \systemname generates offline transformations based on the \pardis API (cf. Figure~\ref{fig:transformation-api}).
Figure~\ref{fig:seq-offline} shows the generated offline transformation for the implementation of the \code{Seq} data-structure using an array. 
This transformation lowers the objects of a \code{Seq} data structure into records with two fields: 1) the underlying array, 2) the current size of the collection.
The nodes corresponding to each method of this data structure are then rewritten to the IR nodes of the implementation body provided in the reflected type.

\begin{figure}[t]
\centering
\begin{lstlisting}
class SeqArrayTransformer extends RuleBasedTransformer{
  rewrite += rule { case SeqNew[T]() =>
    val _maxSize = ("maxSize", true, unit(0))
    val _array = ("array", false, __newArray[T](MAX_BUCKETS))
    record[Seq[T]](_maxSize, _array)
  }
  rewrite += rule { case SetPlusEq[T](self, elem) => 
    self.array.update(self.maxSize, elem)
    self.maxSize_=(self.maxSize.+(unit(1)))
  }
  // Provides access to the fields of the
  // generated record for Seq
  implicit class SeqArrayOps[T](self: Rep[Seq[T]]) {
    def maxSize_=(x: Rep[Int]): Rep[Unit] =   
      fieldSetter(self, "maxSize", x)
    def maxSize: Rep[Int] = 
      fieldGetter[Int](self, "maxSize")
    def array: Rep[Array[T]] = 
      field[Array[T]](self, "array")
  }
}
\end{lstlisting}
\caption{The generated offline transformations by \systemname for \code{Seq} based on arrays.}
\label{fig:seq-offline}
\end{figure}

Many offline transformations require inspecting the generated IR nodes to check their applicability.
In some of these cases, compiler experts can provide annotations to generate the required analysis passes. 
However, in many cases, the analysis requires more features than the ones provided by the existing annotations.
Implementing such analysis passes can be facilitated by using quasi-quotations~\cite{Parreaux:2017:QSR:3136040.3136043,Parreaux:2017:UAS:3177123.3158101,scala-qq}.
More details about the implementation of quasi-quotations and their usages are beyond the scope of this paper.

The aforementioned design provides several advantages over previous work. First,
the \systemname annotation processor uses Scala annotations. This means that
there is no need to provide specific infrastructure for an external DSL, as
opposed to the approach of Stratego/Spoofax~\cite{spoofax}.  Second, developers
can annotate the source code with appropriate annotations, without the need to
port it into another DSL, as opposed to the approach taken in
Forge~\cite{forge}. In other words, developers use the signature of classes and
methods as the meta-data needed for specifying the DSL constructs, whereas in a
system like Forge the DSL developer must use Forge DSL constructs to specify the
constructs of the DSL.  Third, as we aim to give systems developers the ability  to write their systems in plain Scala code, we designed \systemname so that 
developers can place the annotations on
the systems code itself, whereas an approach like Truffle~\cite{truffle-dsl}
focuses on self-optimizing AST interpreters. Thus, the latter annotates the AST
nodes of the language itself.

\subsection{Productivity Evaluation}

\begin{table}[t]
   \centering
   \begin{tabular}{| l || r | r |}
      \hline
      Type & Library & Compiler \\ \hline \hline
      \multicolumn{3}{|c|}{\textbf{Analytical Query Engine}} \\ \hline
     Query Operators & 541 & 3456
\\ \hline
     Monadic Interface & 156 & 407
\\ \hline 
     File Manager & 254 & 291
\\ \hline
Aux. Classes & 100 & 749
\\ \hline \hline
      \multicolumn{3}{|c|}{\textbf{Transactional Query Engine}} \\ \hline
     In-Memory Storage & 45 & 294
\\ \hline 
     Indexing Data-Structures & 69 & 394
\\ \hline 
     Aux. Classes & 58 & 364
\\ \hline \hline
      \multicolumn{3}{|c|}{\textbf{Scala Library}} \\ \hline
Boolean & 18 & 255 
\\ \hline
Int & 85 & 970 
\\ \hline
Seq & 39 & 334 
\\ \hline
Seq Trans. & 176 & 329 
\\ \hline
Array & 39 & 306 
\\ \hline
ArrayBuffer & 52 & 453 
\\ \hline
HashMap & 32 & 259 
\\ \hline
HashMap Trans. & 162 & 305 
\\ \hline
C GLib & 181 & 729 
\\ \hline
Other Classes & 936 & 7007
\\ \hline
\hline
Total & 2943 & 16902\\
\hline
   \end{tabular}
   \vspace{0.5cm}
   \caption{The comparison of LoCs of the \textit{(reflected)} classes of the Scala standard library and a preliminary implementation of two query engines together with the corresponding automatically generated compilation interface.}
   \label{fig:expLocPurg}
\end{table}

We use \systemname to automatically generate the compiler interface for a subset
of the standard Scala library and two database engines: 1) an analytical query engine~\cite{legobase_tods,dblablb}, and 2) a transactional query engine~\cite{dashti_beta}.
Table~\ref{fig:expLocPurg} compares the  number of LoCs\footnote{We used
CLOC~\cite{cloc} to compare the number of LoCs.} of the library classes with the
generated compiler interfaces. We make the following observations.

First, for the Scala standard library classes, the LoCs of the reflected classes
are mentioned in the table. These classes provide the method signatures of their
original classes and are annotated with appropriate effect and algebraic structure 
annotations (Section~\ref{subsec:annotations}). However, in most cases, developers do not need
to provide the implementation of the methods of these classes. As a result, the
compiler interfaces of the Scala standard library classes can be generated with
only tens of LoCs. The exception is the reflected classes responsible for generating offline transformations (e.g., Seq Transformation and HashMap Transformation), where the developer provides the implementation to which every method should be transformed into.

Furthermore, observe that the \code{Int} class contains more LoCs than the many other
standard library classes. This is because each operation of this class encodes
different combinations of arguments in its methods with other numeric
classes (e.g.,\ Int with Double, Int with Float, and so on). Furthermore, the generated
compiler interface of this class is also longer than expected. This is because
the generated compiler code contains the constant-folding optimization
(Section~\ref{subsec:annotations}), which is encoded by \systemname annotations.
In addition, for the query operators of the analytical query engine, the generated compiler interface 
encodes all online partial evaluation processes annotated using the \code{\@inline} annotation. This results in the  partial evaluation of mutable fields,
function inlining, and virtual dispatch removal.

\section{Conclusions}
\label{sec:conclusions}
In this paper, we have presented \systemname, a compiler generator for DSLs embedded in Scala. 
\systemname automatically generates the boilerplate code necessary for building a DSL compiler using the infrastructure provided by existing extensible optimizing compilers. Furthermore, \systemname provides an extensible set of annotations for encoding domain-specific knowledge in different optimizing compilers. 
Finally, we have shown how to extend the \systemname annotations to generate the boilerplate code required for building two different query compilers on top of an extensible optimizing compiler.

\bibliography{refs}

\end{document}